\documentclass[]{spie}  %>>> use for US letter paper
\usepackage[]{graphicx}
\usepackage{graphics}
\usepackage{amssymb}
\usepackage{amsmath}
\usepackage[hyperindex=true,pdftitle={Wavelet-based multifractal analysis of laser biopsy imagery},pdfauthor={Sayantan Ghosh},colorlinks=true,
linkcolor=black,breaklinks=true,citecolor=blue,plainpages=false,
pdfpagelabels]{hyperref}
\usepackage{pstricks}
\usepackage{subfigure}
\usepackage{multirow}
\usepackage{color}
\title{Wavelet-based multifractal analysis of laser biopsy imagery} 

\author{Jaidip Jagtap\supit{a}, Sayantan Ghosh\supit{b}, Prasanta K. Panigrahi\supit{c} and Asima Pradhan\supit{a}
\skiplinehalf
\supit{a}Dept. of Physics, Indian Institute of Technology Kanpur (IIT-K) 208 017, India; \\
\supit{b}Center for Quantum Technology, School of Physics, University of KwaZulu-Natal, Private Bag X54001, Durban 4000, South Africa;\\
\supit{c}Dept. of Physical Sciences, Indian Institute of Science Education and Reasearch Kolkata (IISER-K), P. O. BCKV Campus Main Office, Mohanpur, 741 252, India
}

\authorinfo{Further author information: (Send correspondence to S.G.)\\S.G.: E-mail: 210556397@ukzn.ac.za,say.xen@gmail.com, Telephone: +27 (0)31 230 2809\\  J.J: E-mail: jaidip@iitk.ac.in, P.K.P.: E-mail:pprasanta@iiserk.ac.in and A.P.: E-mail:asima@iitk.ac.in}
 \newcommand{\etAl}{\textit{et al.}}
  
\begin{document} 
  \maketitle 

%%%%%%%%%%%%%%%%%%%%%%%%%%%%%%%%%%%%%%%%%%%%%%%%%%%%%%%%%%%%% 
\begin{abstract}
In this work, we report a wavelet based multi-fractal study of images of dysplastic and neoplastic HE-
stained human cervical tissues captured in the transmission mode when illuminated by a laser light
(He-Ne 632.8nm laser). It is well known that the morphological changes occurring during the
progression of diseases like cancer manifest in their optical properties which can be probed for
differentiating the various stages of cancer. Here, we use the multi-resolution properties of the wavelet
transform to analyze the optical changes. For this, we have used a novel laser imagery technique which
provides us with a composite image of the absorption by the different cellular organelles. As the disease
progresses, due to the growth of new cells, the ratio of the organelle to cellular volume changes
manifesting in the laser imagery of such tissues. In order to develop a metric that can quantify the
changes in such systems, we make use of the wavelet-based fluctuation analysis. The changing self-
similarity during disease progression can be well characterized by the Hurst exponent and the scaling
exponent. Due to the use of the Daubechies' family of wavelet kernels, we can extract polynomial
trends of different orders, which help us characterize the underlying processes effectively. In this study,
we observe that the Hurst exponent decreases as the cancer progresses. This measure could be relatively
used to differentiate between different stages of cancer which could lead to the development of a novel
non-invasive method for cancer detection and characterization.

\end{abstract}

%>>>> Include a list of keywords after the abstract 


%%%%%%%%%%%%%%%%%%%%%%%%%%%%%%%%%%%%%%%%%%%%%%%%%%%%%%%%%%%%%
\section{Introduction}
\label{sec:intro}  % \label{} allows reference to this section
The recent surge in the investigation of non-invasive techniques for cancer detection through fluorescence spectroscopy\cite{alfano1987,kortum1996,ramanujam2000,ghosh2002,biswal2003,ghosh2005}, Raman spectroscopy\cite{haka2005}, elastic scattering spectroscopy\cite{boustany2010,ghosh2011b}, optical gated imaging, optical coherence tomography, diffuse optical tomography, polarization gated imaging \cite{fujimoto2003,schmitt1999,hebden1997,jacques2002}, turbid medium polarimetry \cite{ghosh2011} and phase contrast microscopy\cite{sonibmei2012} are some areas being actively pursued to understand the micro-structure variations through disease progression. It is well known that the elastic scattering spectrum contains rich morphological information about the biological tissue samples due to the inhomogeneity of the constituent organelles' sizes\cite{perelman1998,ghosh2001,gurjar2001,wax2003,drezek2003,graf2005,tuchin2006,ghosh2006,kim2006,choi2007,choi2008,yu2008,kalashnikov2009,boustany2010,ghosh2011}. The angular and wavelength dependence of the elastic scattering spectra have been used to analyze such subtle variations in the morphological changes \cite{perelman1998,wax2003,graf2005,choi2008,kalashnikov2009,ghosh2011b}.
\par
In this work, we perform a Monochromatic Transmission Imaging (MTI) of the tissue samples which provides us with small angle scattering information of the biological samples. Ghosh \etAl\cite{ghosh2011b} have analyzed the angular dependence of the elastic light scattering spectra in order to analyze the changing multi-fractality of the morphological structures in the tissues. Soni \etAl\cite{sonibmei2012}, analyzed the multi-fractality of the refractive index variation captured through phase contrast microscopy where the visible range of the electro-magnetic spectrum was used. The MTI presented in this work is a proof-of-concept study of the multi-fractality of elastic scattering in small forward angles through imaging where the contribution of the larger sized scatters is more pronounced\cite{drezek2003}. It is well known that the elastic light scattering spectra is the power spectrum of the refractive index variation\cite{xu2005,hunter2006,sheppard2007,wu2007,capoglu2009}. Here, we restrict ourselves to a single wavelength in order to observe the refractive index fluctuation at a particular wavelength. 
\par
The use of the Wavelet Based Multi-Fractal De-trended Fluctuation Analysis (WBMFDFA) allows the use of the Multi Resolution Analysis (MRA) capability of the wavelet transforms to isolate trends of different polynomial orders. This is particularly helpful in the context of inhomogeneous size distribution of scatterers in a biological sample. This method has been used in various contexts like determining the multi-fractality in light scattering spectra for pre-cancer detection\cite{ghosh2011b} and studying tissue multi-fractality through phase contrast microscopy\cite{sonibmei2012}. Here we apply this method to explore the possibility of differentiating between various stages of cancer.
\par
This article is organized as follows: In the next section (\ref{sec:EMM}), we give a brief description of the sample preparation and the experimental setup. In the subsequent section (\ref{sec:Theory}), we briefly review the Fourier analysis and the WBMFDFA. In section (\ref{sec:results_d}), we present our observations and discuss the results of the analysis. We conclude with future directions in section (\ref{sec:conclusion}).
\section{Experimental materials and methods}
\label{sec:EMM}
\subsection{Sample preparation}
Hematoxylin and Eosin (HE) stained healthy and histo-pathologically graded neoplastic biopsy samples of human cervical tissues sliced into 4mm $\times$ 6mm (lateral), $\sim$ 5 $\mu$m thick sections were prepared on glass slides for the experiment. The 12 healthy and 22 dysplastic tissue samples were obtained from G. S. V. M. Medical College and Hospital, Kanpur, India. HE staining involves nuclear staining by the application of hemalum (a complex of aluminium ions and haematoxylin) followed by the staining of eosinophilic structures by eosin Y\cite{kiernan2008}. The slide preparation involved standard tissue dehydration, wax embedding and sectioning under a rotary microtome\cite{bancroft2005}. 
\subsection{Experimental setup}
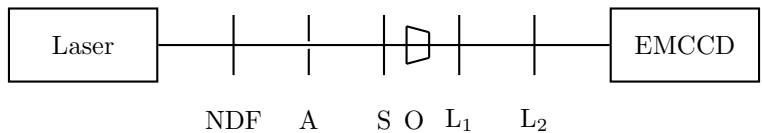
\begin{figure}[htb!]
\centering
\begin{pspicture}(-5,-1.5)(5,1)
%\psgrid
\psline{-}(-3,0)(3,0)
\psframe(-5,-0.5)(-3,0.5) %Laser
\psline{-}(-2,-0.4)(-2,0.4) %NDF
\psline{-}(-1,-0.4)(-1,-0.05)%Aperture bottom
\psline{-}(-1,0.4)(-1,0.05)%Aperture bottom
\psline{-}(0,-0.4)(0,0.4) %Sample
\psline(0.3,-0.25)(0.3,0.25)%Objective
\psline(0.6,-0.17)(0.6,0.17)%Objective
\psline(0.3,-0.25)(0.6,-0.17)%Objective
\psline(0.3,0.25)(0.6,0.17)%Objective
\psline{-}(1,-0.4)(1,0.4)%Lens1
\psline{-}(2,-0.4)(2,0.4)%Lens1
\psframe(5,-0.5)(3,0.5) %EMCCD
\rput(-4,0){Laser}
\rput(4,0){EMCCD}
\rput(-2,-1){NDF}
\rput(-1,-1){A}
\rput(0,-1){S}
\rput(0.4,-1){O}
\rput(1,-1){L$_1$}
\rput(2,-1){L$_2$}
\end{pspicture}
\caption{\label{fig:expsetup} Schematic diagram for the experimental setup. NDF represents a neutral density filter, A is an aperture, S is the sample, O is a 20X objective, L$_1$ and L$_2$ are two collimating lenses. EMCCD is an Electron-Multiplying Charge Coupled Device for collecting the images.}
\end{figure}

\begin{figure}[htb!]
\centering
\subfigure[Healthy epithelium]{
\includegraphics[width=0.4\textwidth]{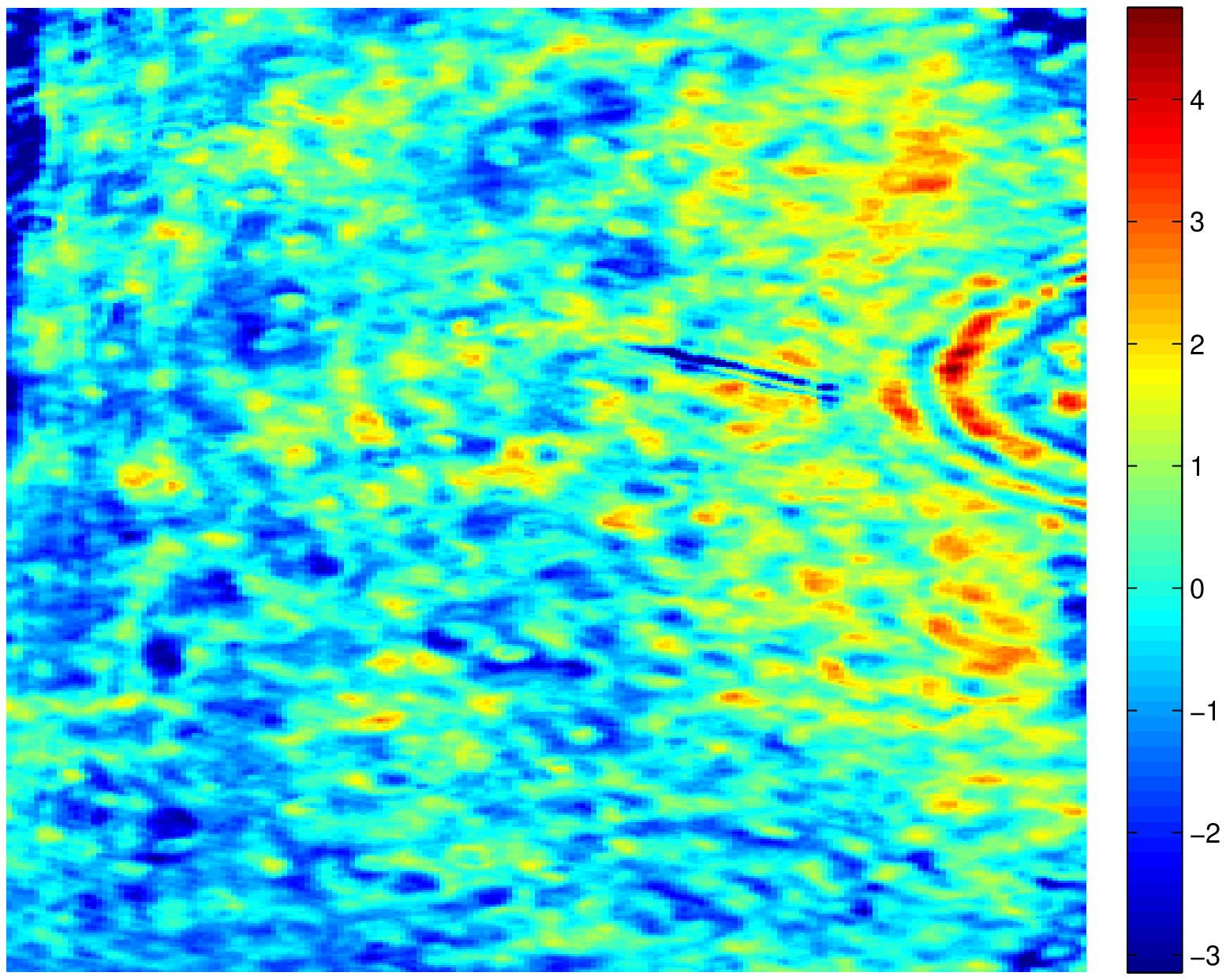}\label{fig:EN_samp}
}
\subfigure[Dysplastic epithelium]{
\includegraphics[width=0.4\textwidth]{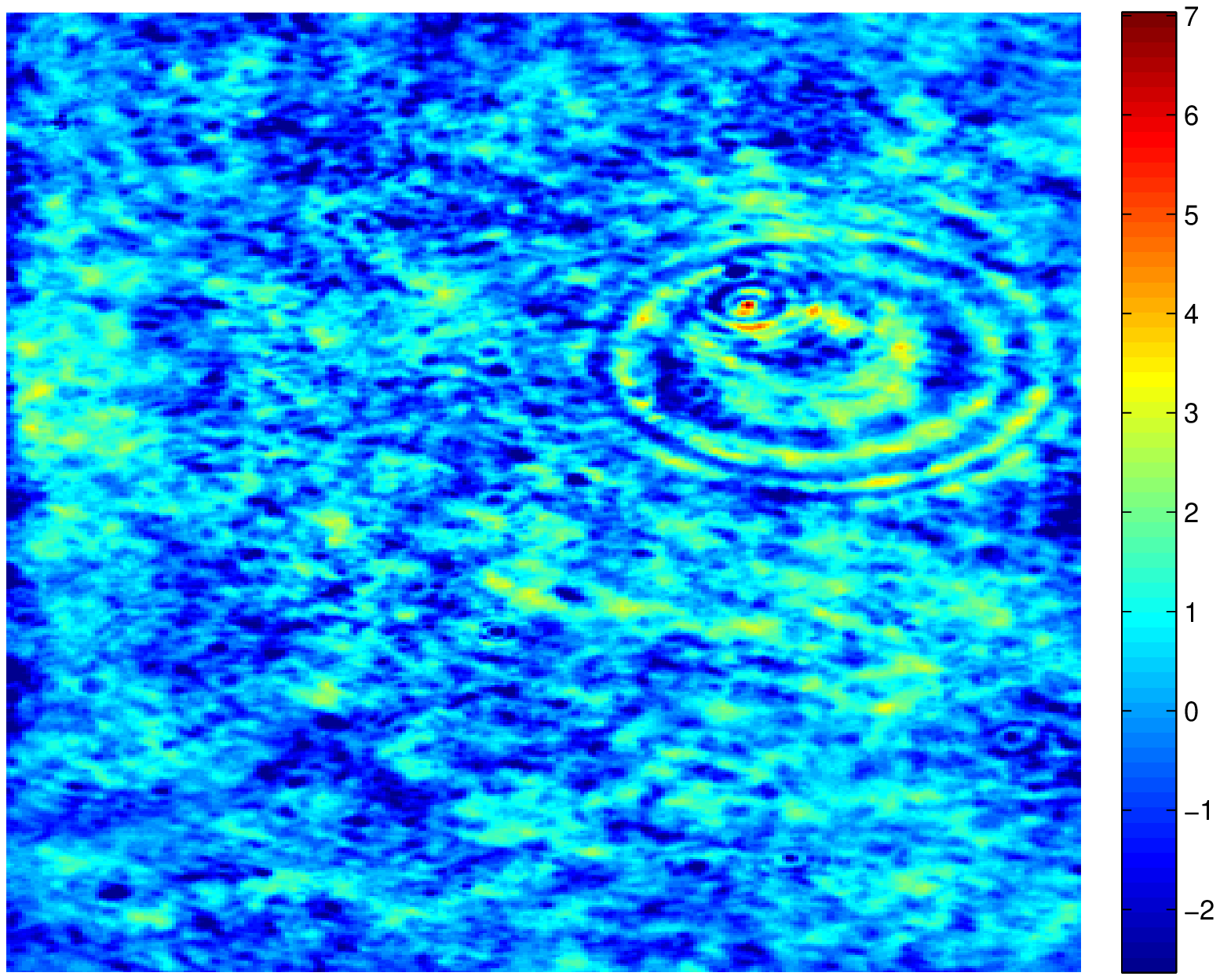}\label{fig:E2_samp}
}
\subfigure[Healthy stroma]{
\includegraphics[width=0.4\textwidth]{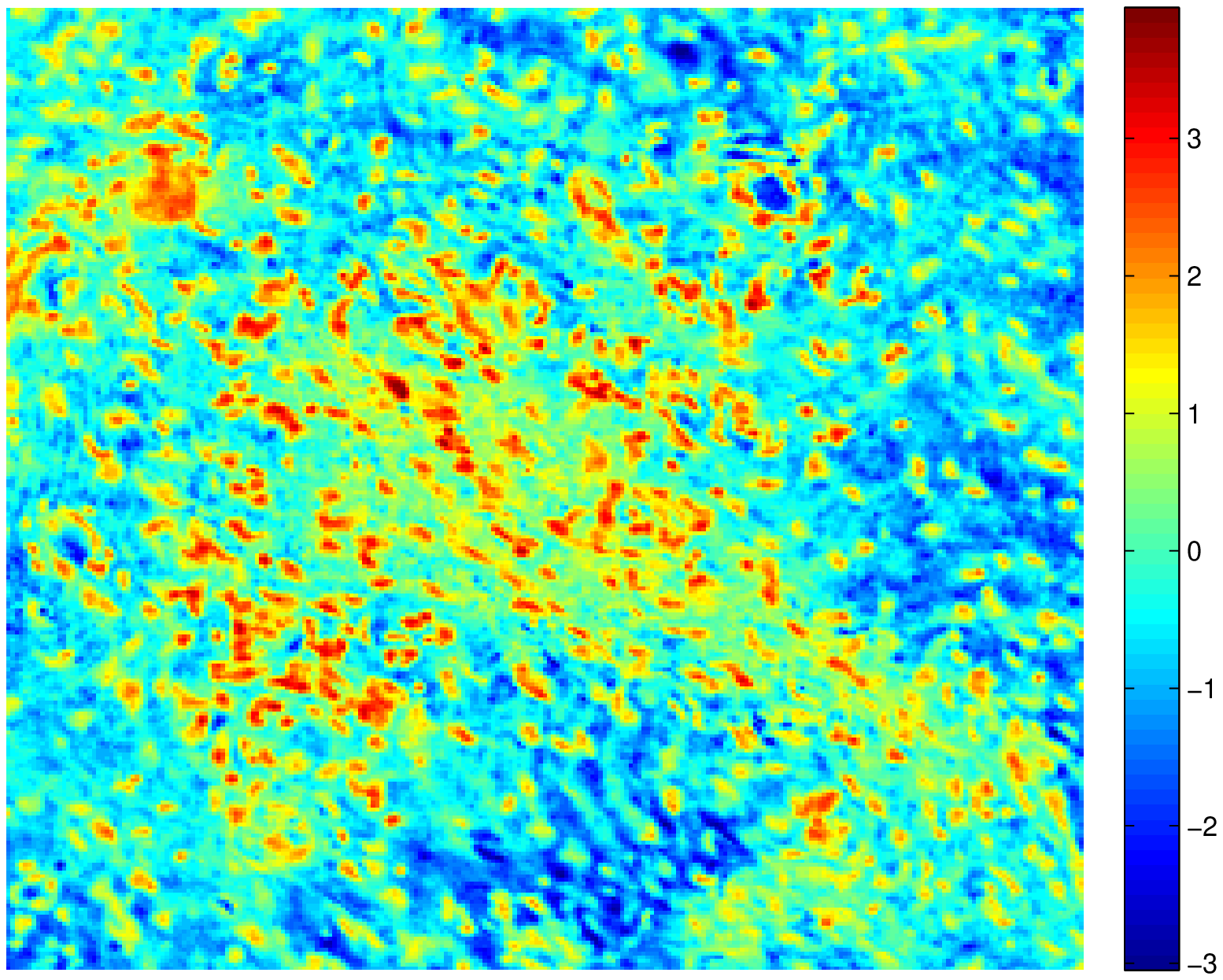}\label{fig:SN_samp}
}
\subfigure[Dysplastic stroma]{
\includegraphics[width=0.4\textwidth]{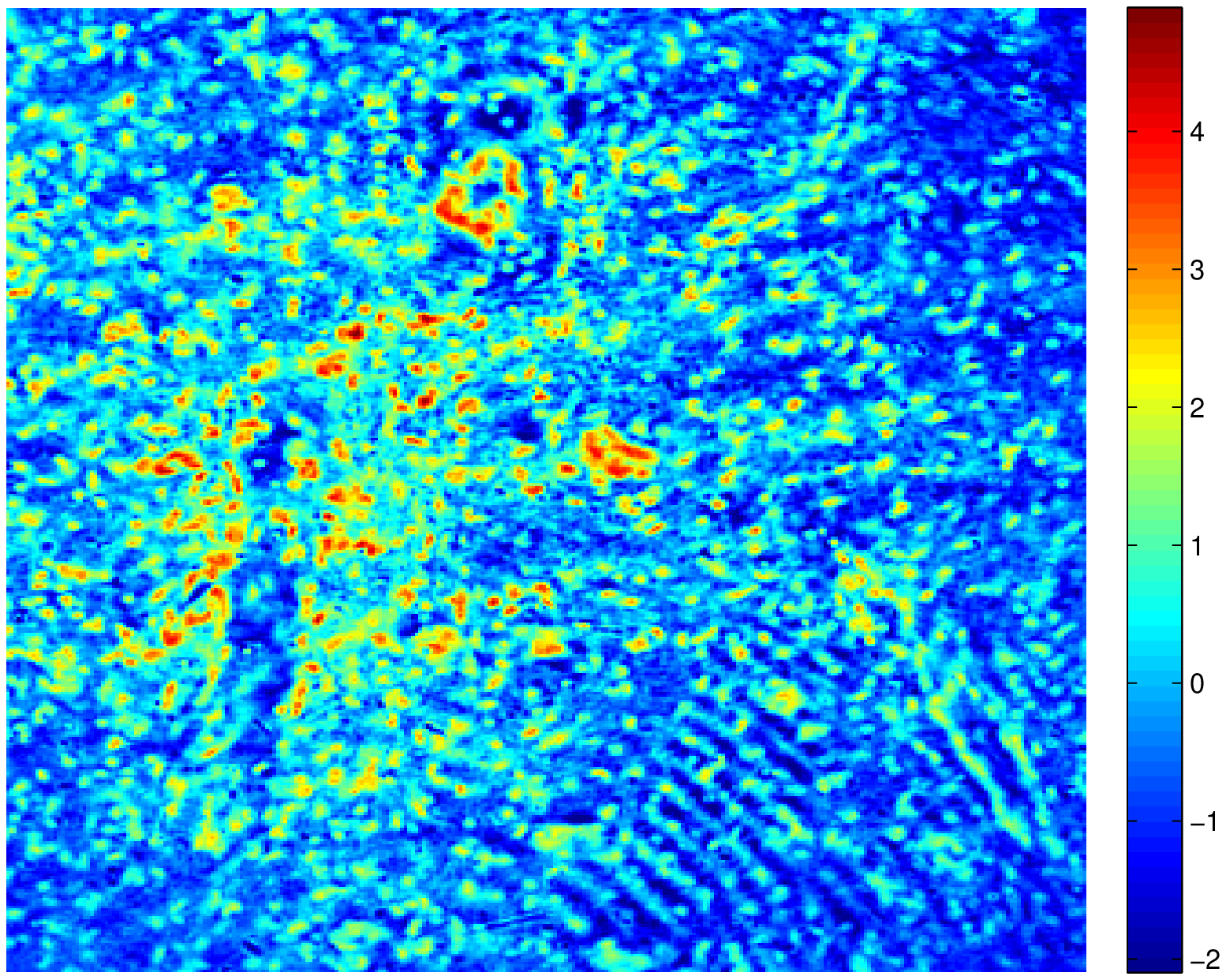}\label{fig:S2_samp}
}
\caption{\label{fig:sample_image}Blank subtracted images of the epithelium for \subref{fig:EN_samp} healthy and \subref{fig:E2_samp} dysplastic tissue samples. Healthy and dysplastic cervical tissue sample of the stroma are shown in \subref{fig:SN_samp} and \subref{fig:S2_samp} respectively.}
\end{figure}

A schematic representation of the experimental setup is shown in Fig. (\ref{fig:expsetup}).
A 632.8nm (output power 5mW) He-Ne laser (Research Electro-Optics Inc., LHRR-0200,USA) masked by a reflective Neutral Density Filter (NDF) (Special Optics Inc., 9-1051, USA; two NDFs were used: a) with optical density 0.5, 31.62\% transmittance and optical density 0.9 with 12.58\% transmittance) was used to illuminate the sample S. An aperture was used to control the beam size to $\sim$ 1mm. The transmitted light was collected through a 20X objective (Labomed LP20X Semi Plan Achro) and collimating lenses on an Electron-Multiplying Charge Coupled Device (Andor iXon3 897, with a pixel size 16$\mu$m $\times$ 16$\mu$m and image size 512 $\times$ 512). This recorded raw image was then subtracted by the image of a blank glass slide to remove the artifacts arising due to glass. The exposure time was kept at 0.4s. The resulting image was then cropped for isolating the epithelium and the stroma. False color images of the epithelium for healthy and dysplastic samples are shown in Fig. (\ref{fig:sample_image}). 
These images were then subjected to fluctuation analysis.
\section{Theory}
\label{sec:Theory}
\subsection{Image unfolding}
The image unfolding is a method to convert two dimensional data into one dimension by horizontally (vertically) concatenating the rows (columns) of the image. For example, a matrix of the form
\begin{equation}
A=\begin{pmatrix}a & b & c\\ d& e &f\end{pmatrix}
\end{equation}
can be unfolded horizontally as
\begin{equation}
\hat{A}_h=(a~b~c~d~e~f)
\end{equation}
and vertically as $\hat{A}_v^T=(a~d~b~e~c~f)^T$ where $^T$ is the matrix transpose. This method has been used to study the multi-fractal behavior of human cervical tissues earlier \cite{sonibmei2012} where the multi-fractality of such tissues were analyzed through phase contrast microscopy.
\subsection{Spectral characterization}
The Fourier power spectrum of a signal $X(\boldsymbol{\xi})$ is given by
\begin{equation}
P(\boldsymbol{\phi})=\left|\hat{X}(\boldsymbol{\phi})\right|^2=\left|\int\limits_{-\infty}^{\infty}X(\boldsymbol{\xi})e^{-\imath\boldsymbol{\xi}\cdot\boldsymbol{\phi}}d\boldsymbol{\xi}\right|^2
\end{equation}
where $\boldsymbol{\phi}$ and $\boldsymbol{\xi}$ are dual spaces of each other and for self-similar processes, the power spectrum is well known to follow a power law
\begin{equation}
P(\boldsymbol{\phi})\sim \boldsymbol{\phi}^{-\alpha}
\end{equation}
where $\alpha$ is called the power law exponent. This is related to the Hurst exponent by $\alpha =2H+1$, $H\in[0,1]$ which is a parameter used to describe the self-similarity of mono-fractal processes and is related to the fractal dimension through $\alpha=2H+1=3-D_f$\cite{hurst1951}. However, the ubiquitousness of multi-fractal processes in nature have been well studied and characterized\cite{stanley1988}. The multi-fractality is characterized by a described by a spectrum of exponents instead of a single exponent. In the next subsection, we will briefly review the multi-fractal analysis through wavelet based fluctuation analysis.
\subsection{Wavelet Based Multi-Fractal De-trended Fluctuation Analysis}
The question of multi-fractal signals has been studied by Stanley and his co workers extensively though the Multi-Fractal De-trended Fluctuation Analysis (MFDFA)\cite{PhysRevE.49.1685}. The Wavelet Based Multi-Fractal De-trended Fluctuation Analysis (WBMFDFA) proposed by Manimaran \textit{et al.} \cite{mani2005,mani2008} used the Multi-Resolution Analysis capability of the wavelet transforms to perform the de-trending of the signals. In this method, we initially make the signal $\bf{X}$ stationary by calculating the log-return series $r(t)=\log(X(t+1))-\log(X(t))$ and normalize it to get the normalized log-return series:
\begin{equation}
R(t)=\frac{r(t)-\langle r(t)\rangle}{\sigma},
\end{equation}
where $\langle r(t)\rangle$ is the time average of the log-return series and $\sigma$ is the standard deviation of $r(t)$. Subsequently, we calculate the profile of the series through
\begin{equation}
Y(t)=\sum\limits_{t'=0}^{t} R(t').
\end{equation}
We use this profile series to extract the fluctuations through discrete wavelet transform. 
\par
The fluctuation extraction involves a wavelet decomposition using the Db4 wavelet which has a support width of 7 and 8 filters\cite{daubechies1992,farge1992}. The profile series can be decomposed as
\begin{equation}
Y(t)=\sum\limits_{j=-\infty}^{\infty}c_j\phi_j(t)+\sum\limits_{i\geq 0}\sum\limits_{j=-\infty}^{\infty}d_{ij}\psi_{ij}(t),
\end{equation}
where, $\psi_{ij}(t)$ is the mother wavelet Db4 and $\phi_j(t)$ is the father wavelet such that it is orthogonal to the mother wavelet. The coefficients $c_j$($d_{ij}$) are called the low pass (high pass) coefficients and capture the trend (fluctuation). The profile is reconstructed at a particular level $j$ by taking only the low pass coefficients to extract the trend at level $j$. This trend is subtract from the profile to obtain the fluctuations at each scale. However, due to the convolution errors, these obtained fluctuations could have edge artifacts which are removed by performing this fluctuation extraction on the reversed profile and taking the average\cite{mani2005,mani2008,ghosh2011a,ghosh2011b}. Then these fluctuations are subdivided in to $n_s$ non-overlapping segments such that $n_s=\lfloor N/s \rfloor$ where the segment length $s$ is related to the wavelet scale $j$ by the number of filter co-efficients for the wavelet used and $N$ is the length of the fluctuations. We obtain the $q^{th}$ order fluctuation function $F_q(s)$ for $q\neq0$ as
\begin{equation}
F_q(s)=\left[\frac{1}{n_s}\sum\limits_{p=1}^{2n_s}\left[F^2(p,s)\right]^{q/2}\right]^{\frac{1}{q}},
\end{equation}
and for $q=0$
\begin{equation}
F_0(s)=\exp\left[\frac{1}{n_s}\sum\limits_{p=1}^{2n_s}\log\left[F^2(p,s)\right]^{q/2}\right]^{\frac{1}{q}},
\end{equation}
The fluctuation function $F_q(s)$ and the window size $s$ are related by
\begin{equation}
F_q(s)\sim s^{h(q)}
\end{equation}
where $h(q)$ is called the generalized Hurst exponent\cite{PhysRevE.49.1685}. However, the dependence of $h(q)$ on $q$ does not make it the ideal parameter for the characterization of multi-fractality \cite{mandelbrot1982} and hence a more sophisticated function called the singularity spectrum is required. 
\par
The singularity spectrum $f(\beta)$ is related to the generalized Hurst exponent by the relations
\begin{eqnarray}
f(\beta)&=&q\left\{\beta-h(q)\right\}+1, ~\mbox{ where,} \\
\beta&=&\frac{d}{dq} \tau(q), ~\mbox{ and}\\
\tau(q)&=&qh(q)-1.
\end{eqnarray}
Here, $\tau(q)$ is the multi-fractal scaling exponent and is defined by the standard partition function based formalism \cite{stanley1988,eke2002} and $f(\beta)$ and $\tau (q)$ are related by a Legendre transform. A quantity $\gamma=\max(\beta)-\min(\beta)$ or the width of the singularity spectrum can be an important parameter for the characterization and differentiation of the multi-fractality of a signal. This parameter has recently been used to characterize the network properties of financial markets\cite{ghosh2011finnet}. We shall use this width of the singularity spectrum to analyze and differentiate between healthy and dysplastic tissue images in the following section. 
\section{Results and discussions}
\label{sec:results_d}
We have shown the false color blank subtracted images of histo-pathologically characterized samples of healthy and dysplastic human cervical tissues in Fig. (\ref{fig:sample_image}). The inhomogeneity of the tissue micro-structure due to the presence of various organelles with various refractive indices can be easily observed in the figures. We use the Fourier analysis and the WBMFDFA in order to probe the refractive index variations associated with the structural changes occurring in the course of disease progression.
\begin{figure}[htb!]
\centering
\subfigure[Healthy sample]{
\includegraphics[width=0.4\textwidth]{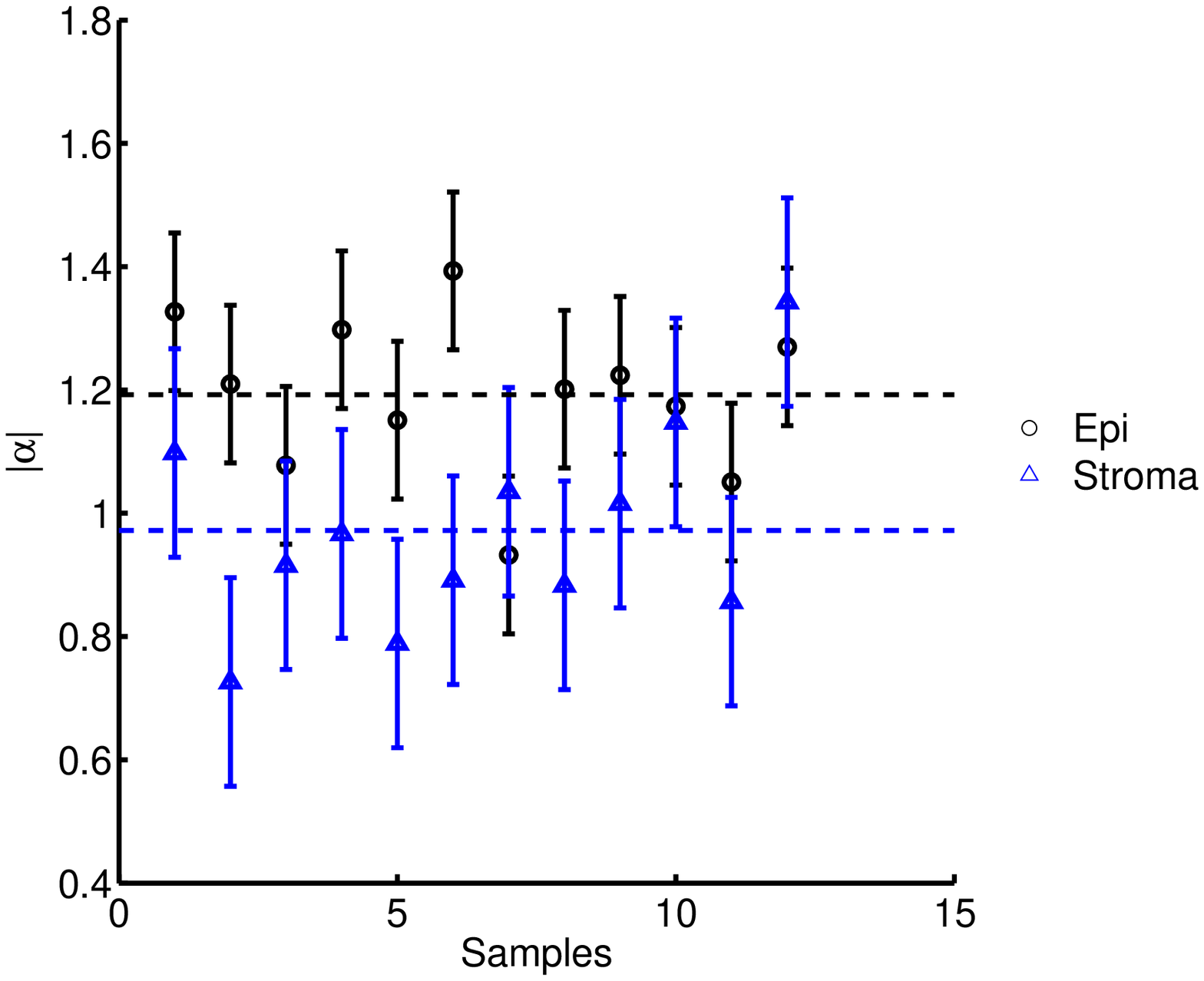}\label{fig:ANE}
}
\subfigure[Dysplastic sample]{
\includegraphics[width=0.4\textwidth]{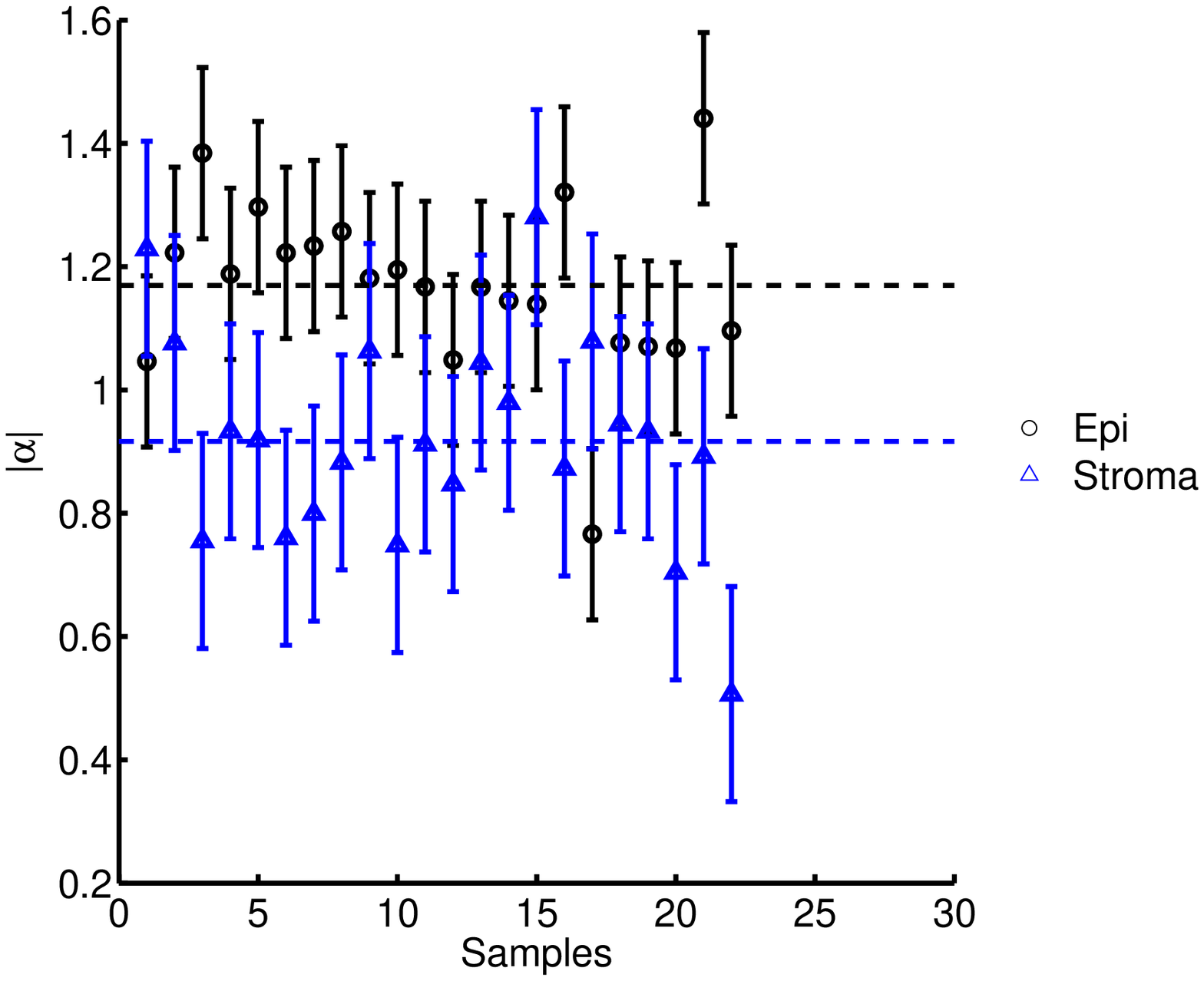}\label{fig:A2E}
}
\caption{\label{fig:AlphaEpi}(Color Online) Power law exponent $\vert \alpha\vert$ for the epithelium for \subref{fig:ANE} healthy and \subref{fig:A2E} dysplastic tissue samples. The mean values for the healthy and cancerous epithelium (stroma) are $\vert\alpha\vert=1.1922\pm0.1279$ ($0.9721\pm0.1692$) and $1.1695\pm0.1389$ ($0.9163\pm0.1745$) respectively.}
\end{figure}
\par
The Fourier power law analysis is based on the mono-fractal hypothesis and proffers the power law exponent $\alpha$ which is related to the Hurst exponent $H$ by $\alpha=2H+1$. The Fourier power in this case is a function of the spatial frequency. The values of the $\alpha$ for the healthy and dysplastic epithelium and stroma are shown in Fig. (\ref{fig:ANE}) and (\ref{fig:A2E}). We observe that the mean absolute $\alpha$ for the epithelium (stroma) is $1.1922\pm0.1279$ ($0.9721\pm0.1692$) for the healthy case while it is $1.1695\pm0.1389$ ($0.9163\pm0.1745$) for the dysplatic case. This power law behavior in the spatial frequency domain can be attributed to the inhomogeneous size distribution of the scatterers in the tissue micro-structure. However, this power law exponent is not independent of the scale as has been observed in the case of the phase contrast microscopy\cite{sonibmei2012} and implies the multi-fractal nature of the scatterer composition of the tissues.
\par
In terms of distinguishing different stages of cancer, as compared to the results obtained in the analysis of the phase contrast microscopic images (where the whole visible spectral region ($400$nm-$800$nm) is used as opposed to this method where a monochromatic image at $632.8$nm is used), the mean value\footnote{The mean value here denotes the ensemble average or the average over all the samples.} of $\alpha$ is lower. Nevertheless, the trend of $\alpha$ for dysplastic samples being lower than that of the healthy samples is corroborated at the $632.8$nm wavelength. As expected, the $\alpha$ for the epithelium $\alpha_e$ is higher than that of the stroma $\alpha_s$. The densely packed structure of the connective fibers in the stroma as compared to the epithelium causes $\alpha_s<\alpha_e$. A comparison of the mean values of $\alpha_s$ and $\alpha_e$ for health and dysplastic tissues is given in table (\ref{tab:tab_param}). To further verify this and to analyze the multi-fractal nature of the tissues, we proceed to study the images through the WBMFDFA.
\begin{figure}[htb!]
\centering
\subfigure[Healthy sample]{
\includegraphics[width=0.4\textwidth]{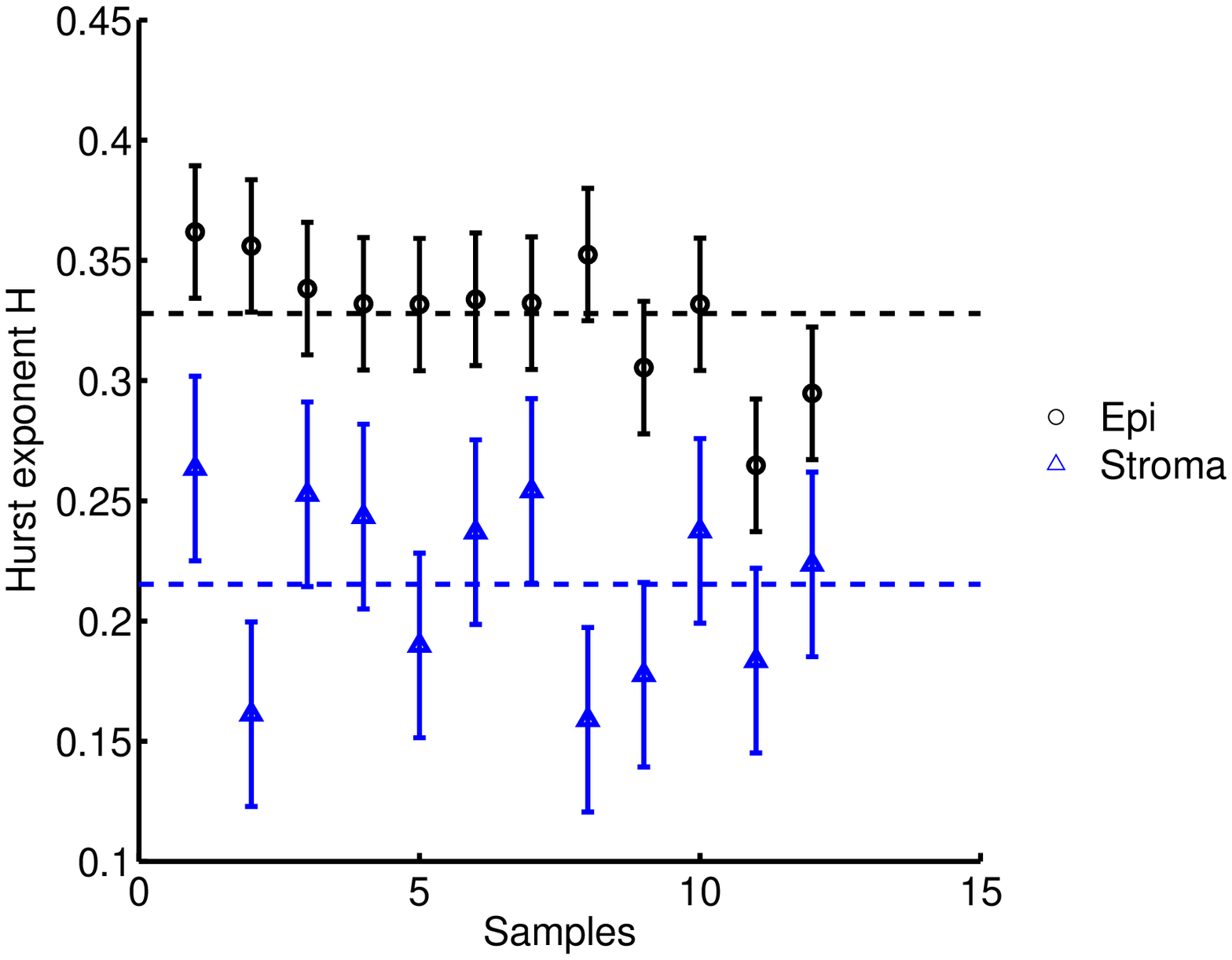}\label{fig:HNE}
}
\subfigure[Dysplastic sample]{
\includegraphics[width=0.4\textwidth]{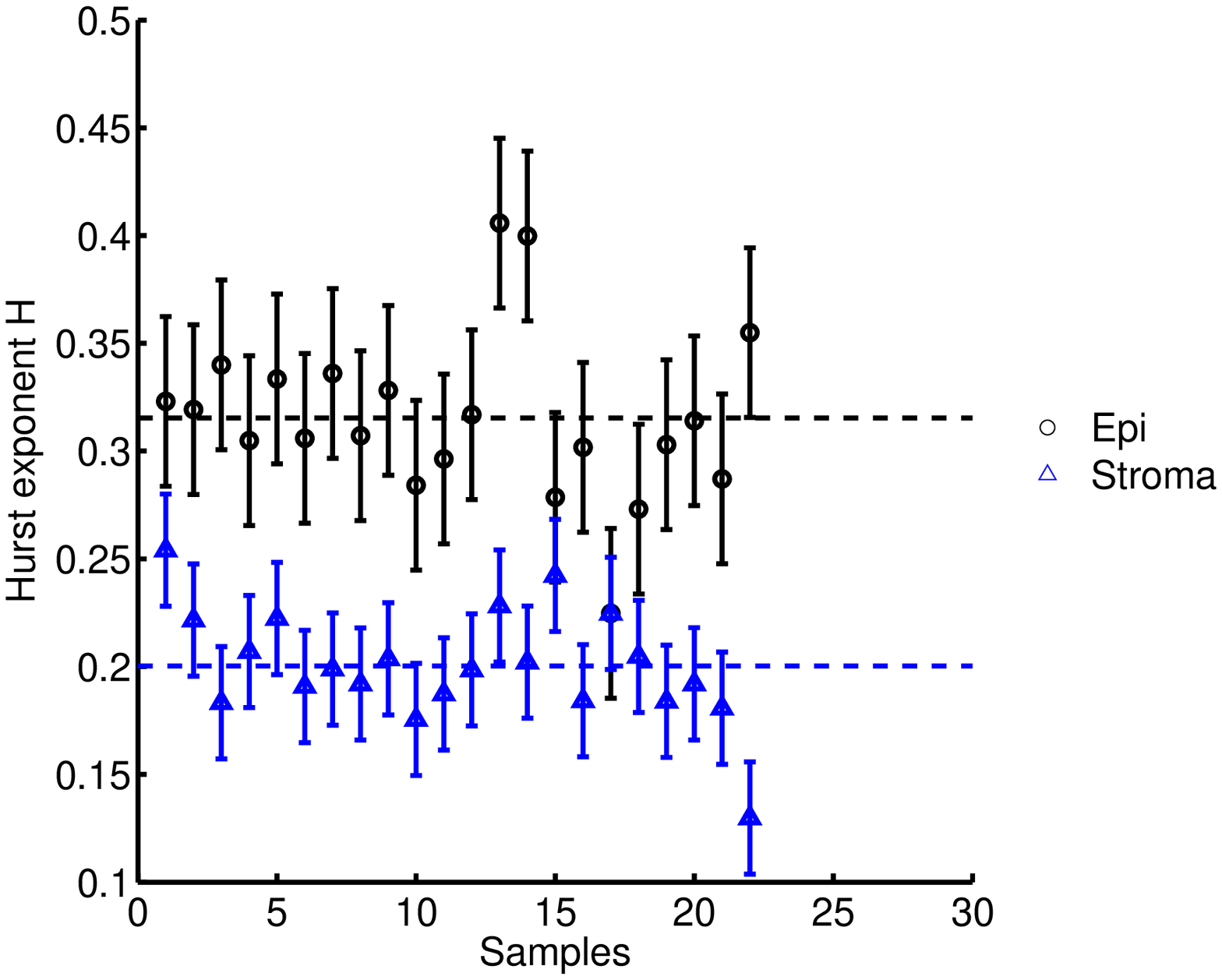}\label{fig:H2E}
}
\caption{\label{fig:Hursttiss}(Color Online) Hurst exponents for \subref{fig:HNE} healthy and \subref{fig:H2E} dysplastic tissue samples. The mean values for healthy and cancerous epithelium (stroma) are $H=0.3279\pm0.0275$ ($0.2152\pm0.0384$) and $0.3154\pm0.0394$ ($0.2003\pm0.0260$) respectively.}
\end{figure}
\par
Figure \ref{fig:Hursttiss} shows the calculated values of the Hurst exponent for healthy and dysplastic tissues in Fig. (\ref{fig:HNE}) and (\ref{fig:H2E}) respectively. The Hurst exponent for the epithelium $H_e$ and that for the stroma $H_s$ are also compared. We can see that while the healthy tissues show a mean $H_e$ of $0.3279\pm0.0275$, the dysplastic tissues display a mean value of $0.3154\pm0.0394$. Similarly, the values of mean $H_s$ are $0.2152\pm0.0384$ and $0.2003\pm0.0260$ for healthy and dysplastic stroma respectively. This is in agreement with the results of the Fourier analysis based power law calculations. As observed earlier, the $H_e>H_s$. The comparative values of $H_s$ and $H_e$ are given in table (\ref{tab:tab_param}).
\begin{figure}[htb!]
\centering
\subfigure[Healthy sample]{
\includegraphics[width=0.4\textwidth]{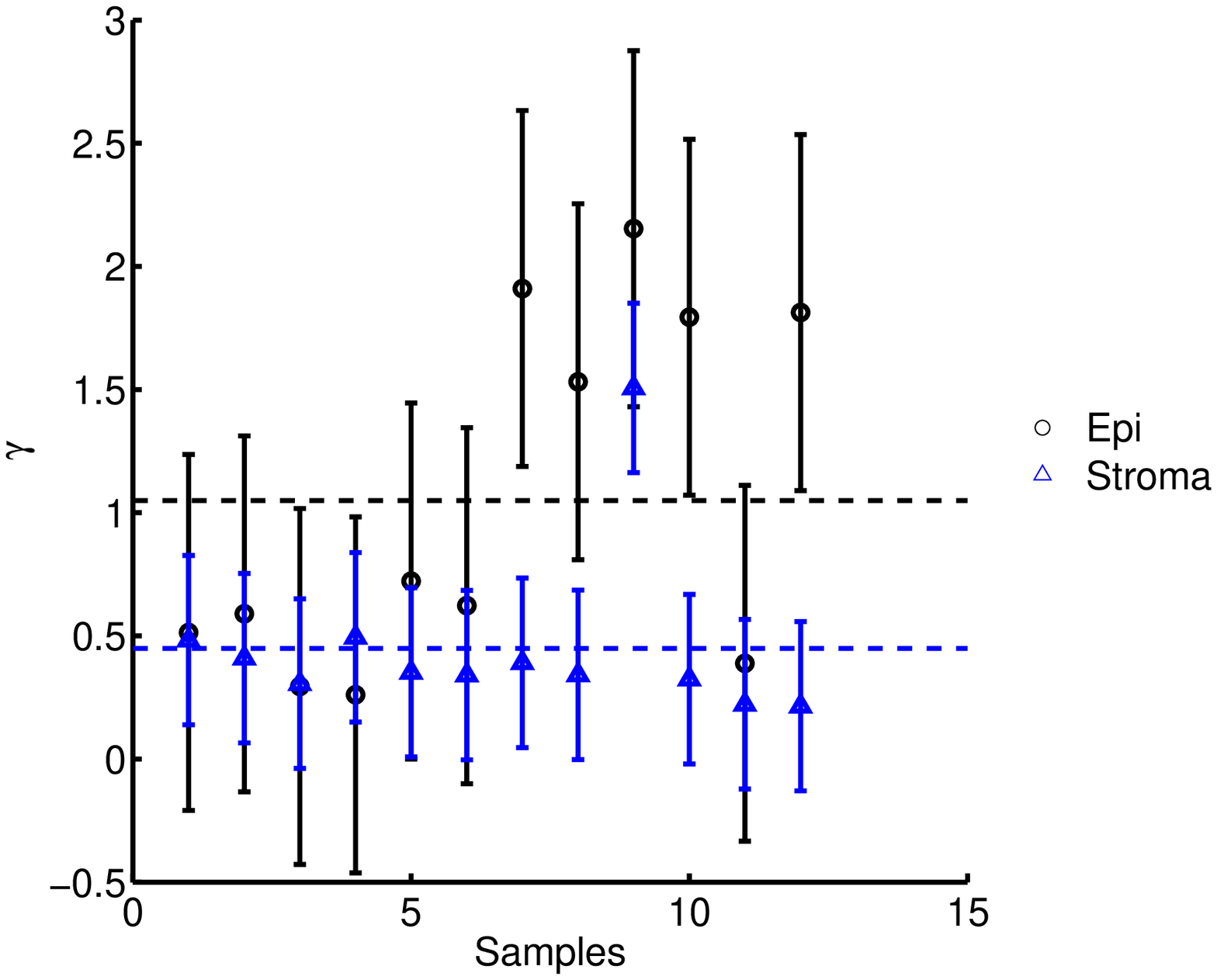}\label{fig:GNE}
}
\subfigure[Dysplastic sample]{
\includegraphics[width=0.4\textwidth]{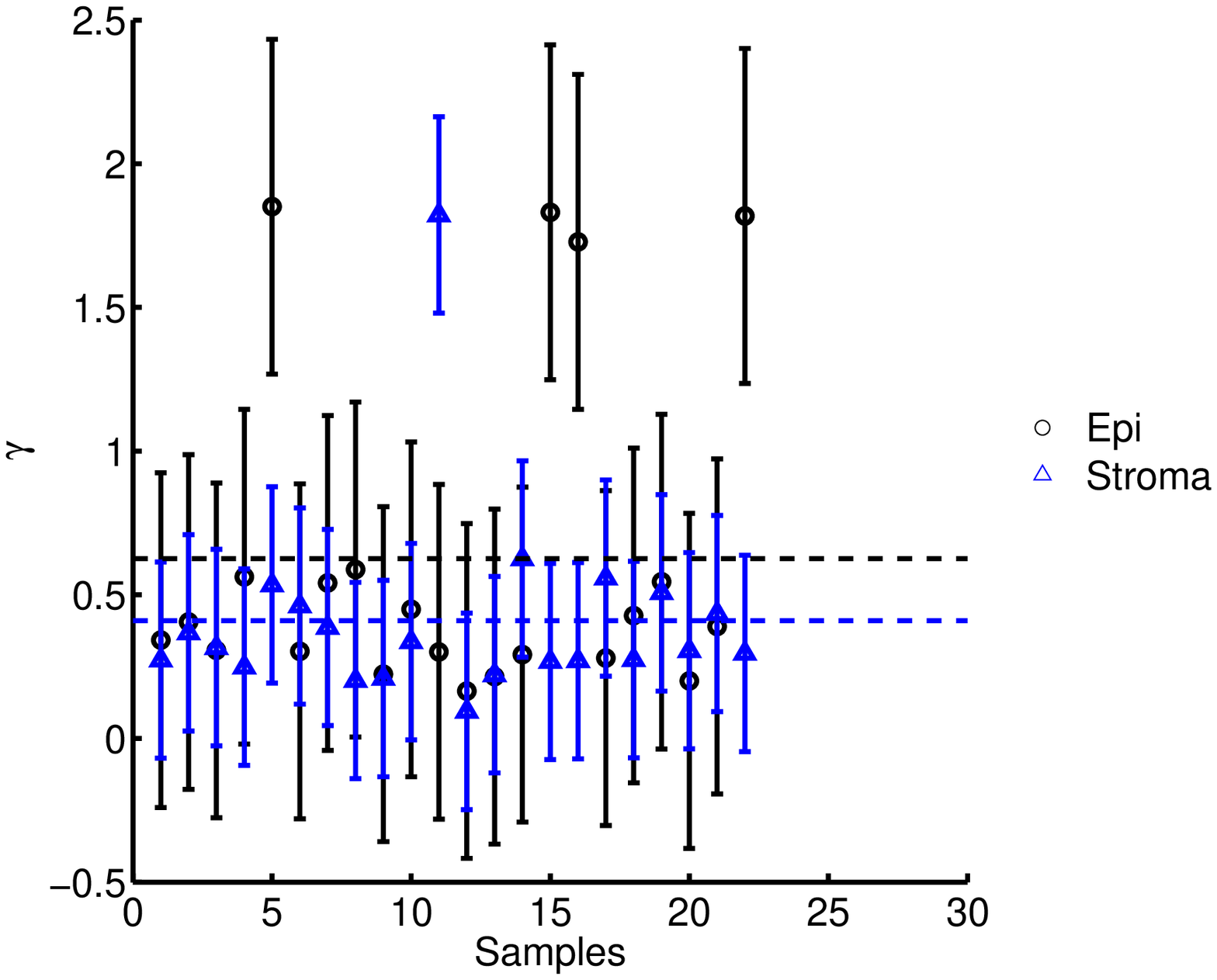}\label{fig:G2E}
}
\caption{\label{fig:GammaEpi}(Color Online) Width of the singularity spectrum $\gamma$ for \subref{fig:GNE} healthy and \subref{fig:G2E} dysplastic tissue samples. The mean values for healthy and cancerous epithelium (stroma) are $\gamma=1.0493\pm0.7226$ ($0.4486\pm0.3440$) and $0.6257\pm0.5827$ ($0.4095\pm0.3416$) respectively.}
\end{figure}
\par
We mentioned in the earlier section that multi-fractal signals require the singularity spectrum for characterization and that the width of the singularity spectrum $\gamma$ is a parameter used to describe multi-fractality of the data under consideration. In Fig. (\ref{fig:GammaEpi}), the $\gamma_s$ and $\gamma_e$ for healthy \ref{fig:GNE} and dysplastic \ref{fig:G2E} tissues is shown. The mean $\gamma_e$ and $\gamma_s$ are found to be $1.0493\pm0.7226$ and $0.6257\pm0.5827$ respectively. As compared to the differences between $\alpha_e$ and $\alpha_s$ and $H_e$ and $H_s$; the difference between $\gamma_e$ and $\gamma_s$ is  more pronounced. Still, the trend of $\gamma_e>\gamma_s$ is followed, also implying the higher multi-fractality of the epithelium as compared to the stroma.
\begin{table}[htb!]
\centering
\begin{tabular}{c c c c c}
\hline
\multirow{2}{*}{Parameter}&\multicolumn{2}{c}{Epithelium}&\multicolumn{2}{c}{Stroma}\\
\cline{2-5}
&Healthy&Dysplastic&Healthy&Dysplastic\\
\hline
$[\alpha]$&$1.1922\pm0.1279$&$1.1695\pm0.1389$&$0.9721\pm0.1692$&$0.9163\pm0.1745$\\
$[H]$&$0.3279\pm0.0275$&$0.3154\pm0.0394$&$0.2152\pm0.0384$&$0.2003\pm0.0260$\\
$[\gamma]$&$1.0493\pm0.7226$&$0.6257\pm0.5827$&$0.4486\pm0.3440$&$0.4095\pm0.3416$\\
\hline
\end{tabular}
\caption{\label{tab:tab_param} Comparison of the mean parameters calculated through the Fourier and the WBMFDFA. $[\cdots]$ represents the ensemble average over the different samples of healthy and dysplastic tissues. The parameters have been averaged over $12$ and $22$ samples of healthy and dysplastic cervical tissues respectively.}
\end{table}
\section{conclusion}
\label{sec:conclusion}
In conclusion, we have compared the normal and dysplastic human cervical epithelium and stroma in an effort to probe the differences between their multi-fractality towards developing a possible non-invasive optical technique for cancer detection. The use of monochromatic imagery for this purpose provides us information about the micro-structural changes in the tissues associated with disease progression.
\par
We observe that the multi-fractality of tissues decreases with progression of cancer. Though the $\alpha$ or $\gamma$ are not drastically different, the trend of decreasing multi-fractality through the progression is sufficient to quantify the stage of cancer. For example, the decreasing $\alpha$ shows that the as the cancer progresses, the tissue micro-structure goes from $~1/f$ towards a flatter power spectrum. Similarly, the Hurst exponents show that with the increase in the cancer grades, the structural organization goes from a seemingly more random behavior to a spatially long term correlated behavior. The decreasing $\gamma$ implies the decrease in multi-fractality with the disease progression.
\par
This work is a proof of concept about the exploration of monochromatic laser imagery to understand the variations of the tissue micro-structure when observed at a particular wavelength as compared to phase contrast microscopy where the whole visible range is used. However, we believe that due to the presence of various fluorescing enzymes in the cancerous tissues like NAD$^{\mbox{+}}$, the response of the tissues to different wavelengths would be different and a more extensive study of the same is being undertaken which will be reported soon. A notable feature of this study is its candidature for \textit{in-vivo} examination and characterization of the MTI which is also under investigation currently.

\end{document}